\documentclass{PoS}
\usepackage{subfigure}
\title{Meson spectra from overlap fermion on domain wall gauge configurations}

\ShortTitle{Meson spectra from overlap fermion}
\author{ {\speaker{N. Mathur}}$^a$, A. Alexandru$^b$, Y. Chen$^c$, T. Doi$^d$, S.J. Dong$^e$, T. Draper$^e$, M. Gong$^e$,
\mbox{F.X. Lee$^b$,} A. Li$^f$, K.-F. Liu$^e$, T. Streuer$^g$ and J.B. Zhang$^h$}

\author{{\hspace*{2.0in}}($\chi$QCD Collaboration)\\
\llap{$^a$}Department of Theoretical Physics, Tata Institute of Fundamental Research, Homi Bhabha Road, Mumbai 400005, India\\
\llap{$^b$}Dept. of Physics, George Washington University, Washington, DC 20052, USA\\
\llap{$^c$}Institute of High Energy Physics, Chinese Academy of Science, Beijing 100049, China\\
\llap{$^d$}Graduate School of Pure and Applied Science, University of Tsukuba, Tsukuba, Ibaraki 305-8571, Japan\\
\llap{$^e$}Dept. of Physics and Astronomy, University of Kentucky, Lexington, KY 40506, USA\\
\llap{$^f$}Dept. of Physics, Duke University, Durham, NC 27708, USA\\
\llap{$^g$}Institute for Theoretical Physics, University of Regensburg, 93040 Regensburg, Germany\\
\llap{$^h$}ZIMP and Dept. of Physics, Zhejiang University, Hangzhou, Zhejiang 310027, China\\
        E-mail: \email{nilmani@theory.tifr.res.in}}

\abstract{We report meson spectra obtained by using valence overlap fermion propagators generated on a background
of 2+1 flavor domain wall fermion gauge configurations on $16^3 \times 32$, $24^3 \times 64$ and $32^3 \times 64$
lattices. We use many-to-all correlators with $Z_3$ grid source and low eigenmode substitution which is efficient
in reducing errors for the hadron correlators.
The preliminary results on meson spectrum, $a_0$ correlators, and charmonium hyperfine splitting for three sea quark masses are reported here.}

\FullConference{The XXVIII International Symposium on Lattice Field Theory, Lattice2010\\
		June 14-19, 2010\\
		Villasimius, Italy}

\begin{document}
\vspace*{-0.5in}
\section{Introduction}
During the last few years RBC and UKQCD collaborations have generated a
large set of configurations with 2+1 flavor dynamical domain wall
fermions (DWF) %with Iwasaki gauge action 
on several lattices with
pion mass as low as 300 MeV and volume large enough for mesons
($m_{\pi}L$ > 4)~\cite{RBC-UKQCD08}. Reducing sea quark masses to
lower values requires substantial computational effort.  A
possible expedient approach toward unquenched QCD simulation with
chiral fermions could be to use overlap fermions as the valence quark
on those domain wall gauge configurations. In view of the fact that
the overlap fermion has additional desirable features, such as 
incorporation of multi-mass algorithms and precise evaluation of the
matrix sign function, which one can take advantage of in order to
improve chiral symmetry as well as the quality of the numerical
results.

The use of overlap valence on domain wall fermion gauge configurations is a mixed action approach which is associated with both mixed action and partially quenched disadvantages. However, the mixed
action partially quenched chiral perturbation theory (MAPQ$\chi$PT)
has been developed for Ginsparg-Wilson fermion on Wilson
sea~\cite{brs03} and staggered sea~\cite{bbr05}, and has been worked
out for many hadronic quantities to next-to-leading order (NLO), such
as pseudoscalar masses and decay constants~\cite{brs03,bbr05},
isovector scalar $a_0$ correlator~\cite{gis05,pre06,wlo09},
heavy-light decay constants~\cite{ab06}, and baryon
masses~\cite{tib05,wlo09}. Using similar MAPQ$\chi$PT for two
different chiral fermions, it is possible to extract reliable physical
quantities such as masses and decay constants.  Since both fermions
used here are chiral fermions, $\Delta_{mix}$, the
low energy constant representing $\mathcal{O}(a^2)$ discretization
dependence, is smaller than other mixed action
formulations~\cite{big-paper}. 
We shall also adopt the deflation method 
for overlap formulation, as will be mentioned later, 
to speed up inversion.
The overlap formulation also incorporates
multi-mass algorithm which enables calculation of multiple
quark propagators covering the range from very light quarks to the
charm on these sets of DWF lattices. This makes it possible to include
the charm quark for calculations of charmonium and charmed-light
mesons using the same fermion formulation for the charm and light
quarks~\cite{dl09}. With this formalism, while it is possible to get
insight of light quark behaviors of many physical quantities, rich
phenomenology involving charm mesons and baryons can also be studied in
the same lattice formulation.

\vspace*{-0.1in}
\section{Formalism : Overlap with deflation, HYP smearing, $Z_3$ grid source and low mode substitution}
The formalism for our calculation was detailed in ref.~\cite{big-paper}. Here we outline that briefly. It has been shown~\cite{wil07} that the projection of low
eigenmodes from the Dirac operator can speed up inversion of fermion
matrices and this procedure of deflation has been applied to both
hermitian~\cite{ehn99,dll00} and non-hermitian~\cite{mw07} systems as
well as to hermitian system with multiple right-hand
sides~\cite{so07}.  In addition to faster inversion one can also
substitute exact low eigenmodes in the noise estimation such as in
quark loops~\cite{nel01} and all-to-all
correlators~\cite{fjo05,kfh07} to reduce errors in two and
three-point correlators of mesons.

One can get the eigenvectors of the massive overlap Dirac operator by calculating the same for the massless Dirac operator $D_{ov}$. Since $D_{ov}^{\dagger} = \gamma_5 D_{ov} \gamma_5$, and $[D_{ov}D_{ov}^{\dagger}, \gamma_5] =0$
one first uses Arnoldi algorithm to search for eigenmodes of $D_{ov}D_{ov}^{\dagger}$
with real eigenvalues $|\lambda_i|^2$ which are doubly degenerate with opposite chirality.
Eigenmodes of $D_{ov}$ can then be obtained by diagonalizing the two chiral modes in $D_{ov}$.
\newpage
\vspace*{-0.3in}
Smearing the gauge links suitably could deplete the
density of the lowest eigenvalues in $H_W$ and by using HYP smearing we observe that the lowest eigenvalue with HYP smearing after deflation with
100 to 200 eigenmodes is about 3 times larger than those without HYP smearing.
In Fig.1 we have plotted the spectra of the lowest 200 (400) eigenvalues for kernel in the
           inner (outer) loop of the overlap fermion for a $32^3 \times 64$ configuration with
           $m_l = 0.004$.
\begin{figure}
\vspace{-0.2in}
\begin{center}
\includegraphics[width=0.47\textwidth,height=0.28\textwidth,clip=true]{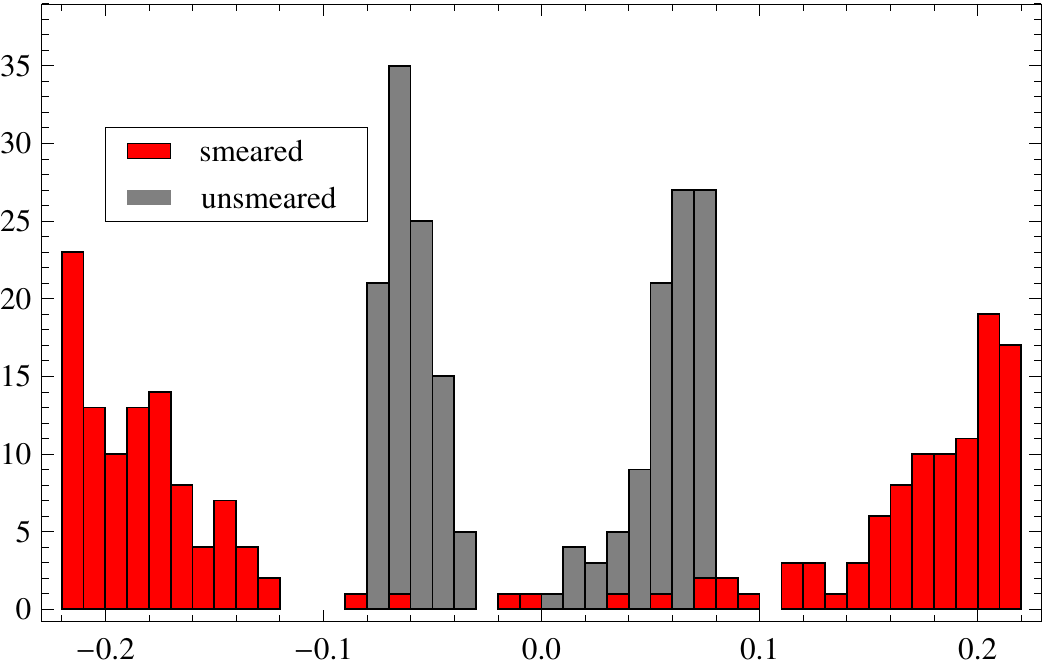}
\hspace*{0.2in}\includegraphics[width=0.47\textwidth,height=0.28\textwidth,clip=true]{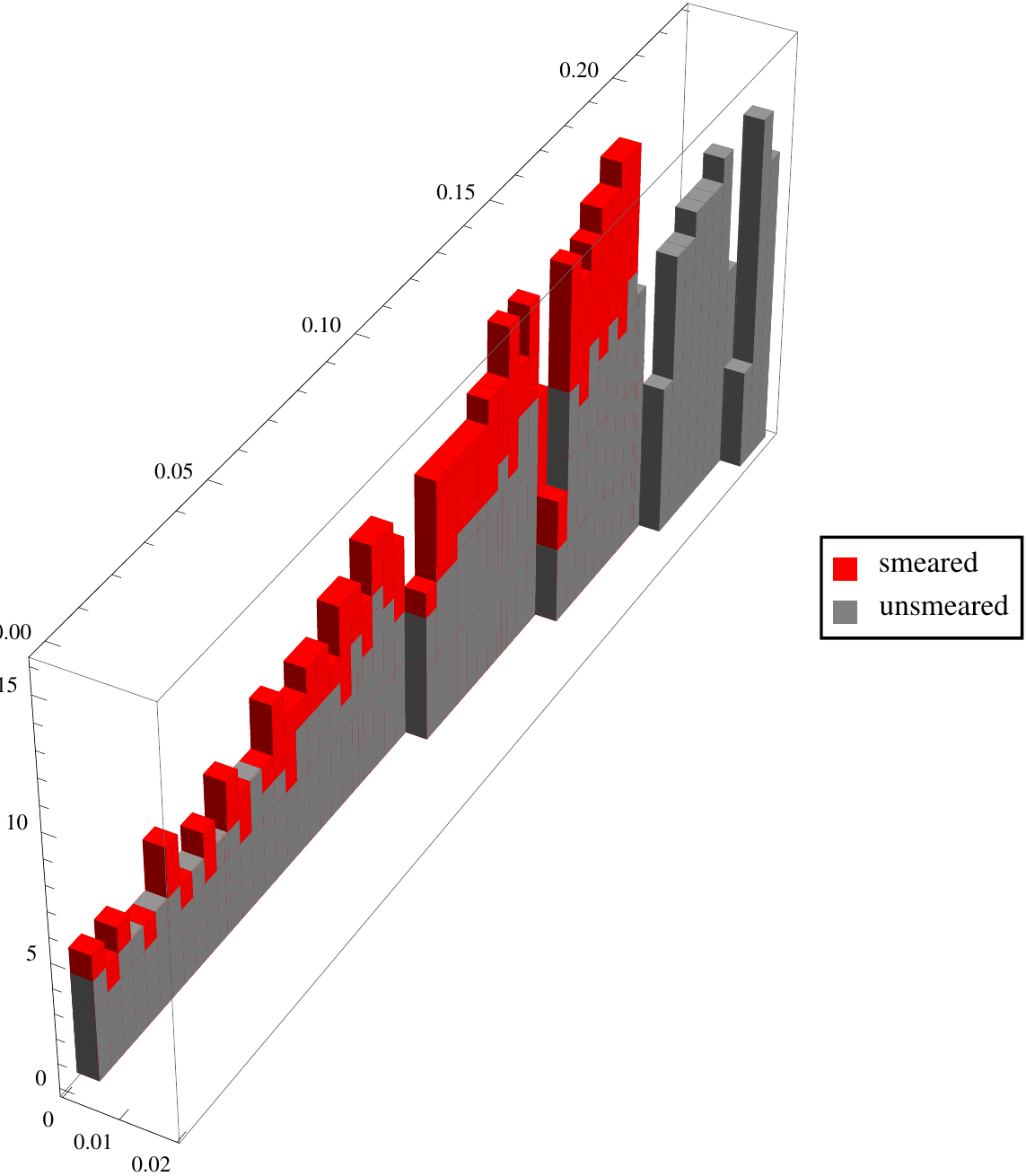}
\end{center}
\vspace{-0.17in}
\caption{(left (right)) The lowest 200 (400) eigenvalues for kernel in the
           inner (outer) loop of the overlap fermion for a $32^3 \times 64$ configuration with
           $m_l = 0.004$.}
\end{figure}
In Table~\ref{comparison} we sum up our findings for speed up using HYP smearing and deflation in detail along with overhead of producing eigenmodes.
\vspace*{-0.2in}
\begin{table}[ht]
\caption{A comparison of speedup of inversion with HYP smearing (S) and deflation (D) of the outer loop.
The inner and outer iteration
numbers are for one propagator with 12 columns of color-spin. The speedup refers to that between the case of
$D+S$ vs the one with neither $D$ nor $S$.
\label{comparison}}
{\footnotesize
\begin{center}
\begin{tabular}{|l|c|ccc|ccc|ccc|}
  \hline
  \multicolumn{1}{|c}{} &\multicolumn{4}{c|}{$16^3\times 32$}
  & \multicolumn{3}{c|} {$24^3 \times 64$} & \multicolumn{3}{c|} {$32^3 \times 64$}\\
  \hline \hline
     & residual & w/o D & D & D+S  & w/o D & D & D+S & w/o D & D & D+S \\
  \hline
  $\rm{H_W}$ eigenmodes & $10^{-14}$ & 100 & 100 & 100 &  400 & 400 & 400 & 200 & 200 & 200 \\
  $D_{ov}$ eigenmodes & $10^{-8}$ & 0 & 200 & 200 &  0 & 200 & 200 & 0 & 400 & 400 \\
  Inner iteration & $10^{-11}$ & 340 & 321 & 108  & 344 & 341 & 107 &  309 & 281 & 101\\
  Outer iteration & $10^{-8}$ & 627 & 72 & 85  & 2931 & 147  & 184 & 4028 & 132 & 156  \\
  \hline
  Speedup    &     &    &  &  23 &     &     & 51 &      &  & 79    \\
    Overhead   &     &    &  \multicolumn{2}{r|} {4.5  propagators}  &  & \multicolumn{2}{r|} {4.9 propagators} &  &
\multicolumn{2}{r|} {7.9 propagators} \\
  \hline
\end{tabular}
\end{center}
}
 \end{table}
%\begin{figure}
%\begin{center}
%\includegraphics[width=0.47\textwidth,height=0.28\textwidth,clip=true]{PS_888_mass4_rel.pdf}
%\includegraphics[width=0.47\textwidth,height=0.28\textwidth,clip=true]{PS_888_mass4.pdf}
%\end{center}
%\vspace{-0.15in}
%\caption{(a) $(m_{\pi}a)^2$ is plotted as a function of $ma$ for the $32^3\times 64$ lattice with quark masses $ma$ = [0.00275, 0.3]. (b) $(m_{\pi}a)^2/ma$ is plotted vs $ma$ showing chiral logarithm for the same range of quark masses.}
%\end{figure}
\begin{figure}
\begin{center}
\includegraphics[width=0.47\textwidth,height=0.28\textwidth,clip=true]{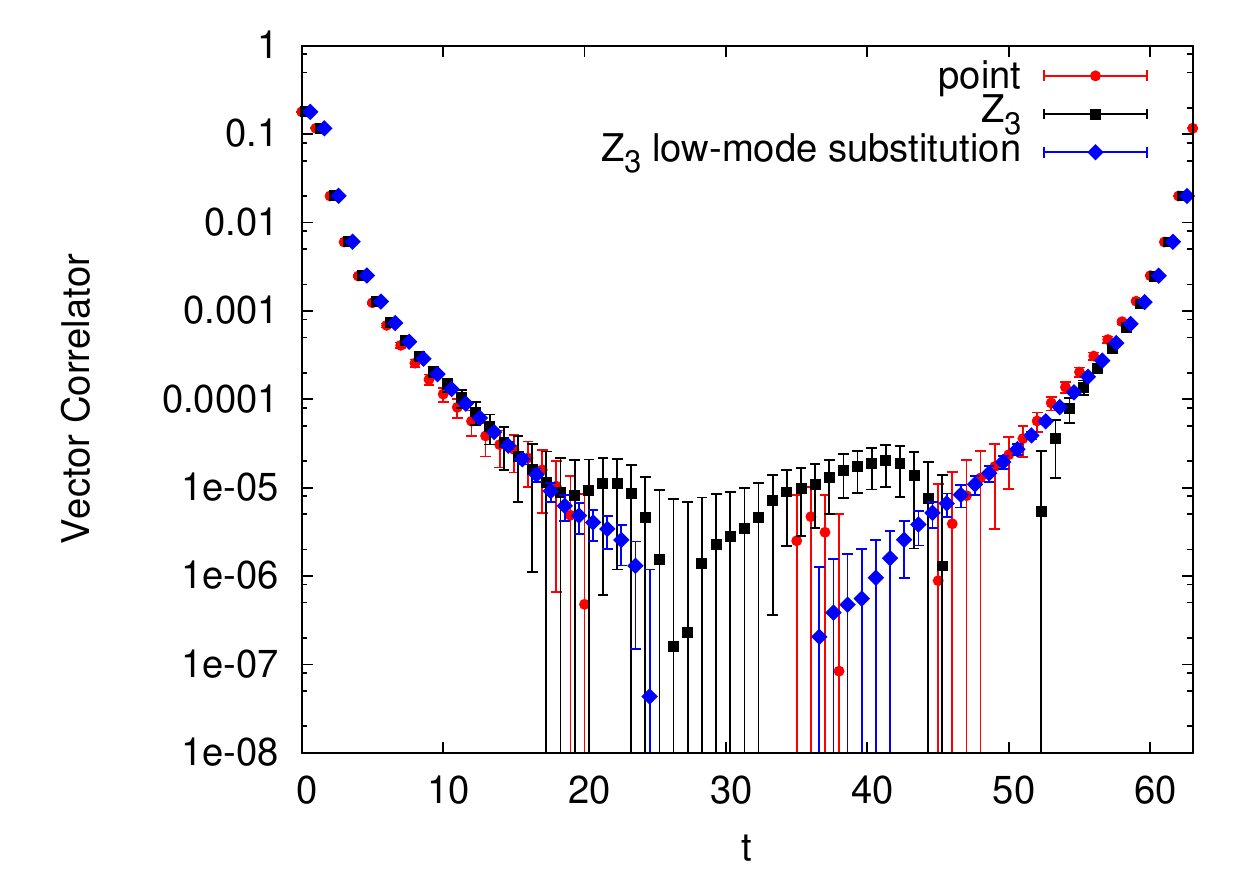}
\includegraphics[width=0.47\textwidth,height=0.28\textwidth,clip=true]{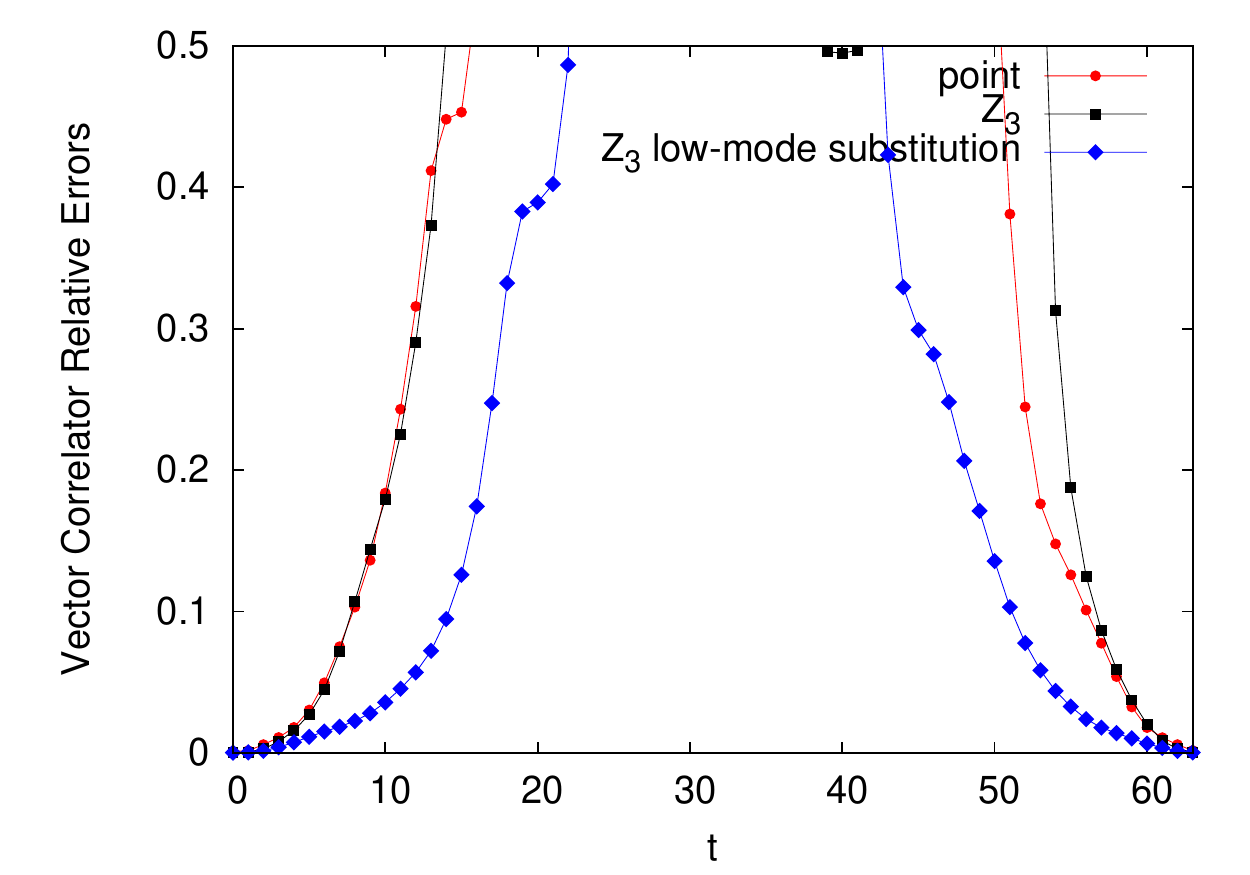}
\end{center}
\vspace{-0.25in}
\caption{(left) The vector correlator for the $32^3\times 64$ lattice at $m_{\pi} \sim 200$ MeV for three different choices. (right) The respective relative errors.}
\end{figure}
\begin{figure}
\vspace{-0.2in}
\begin{center}
\includegraphics[width=0.47\textwidth,height=0.28\textwidth,clip=true]{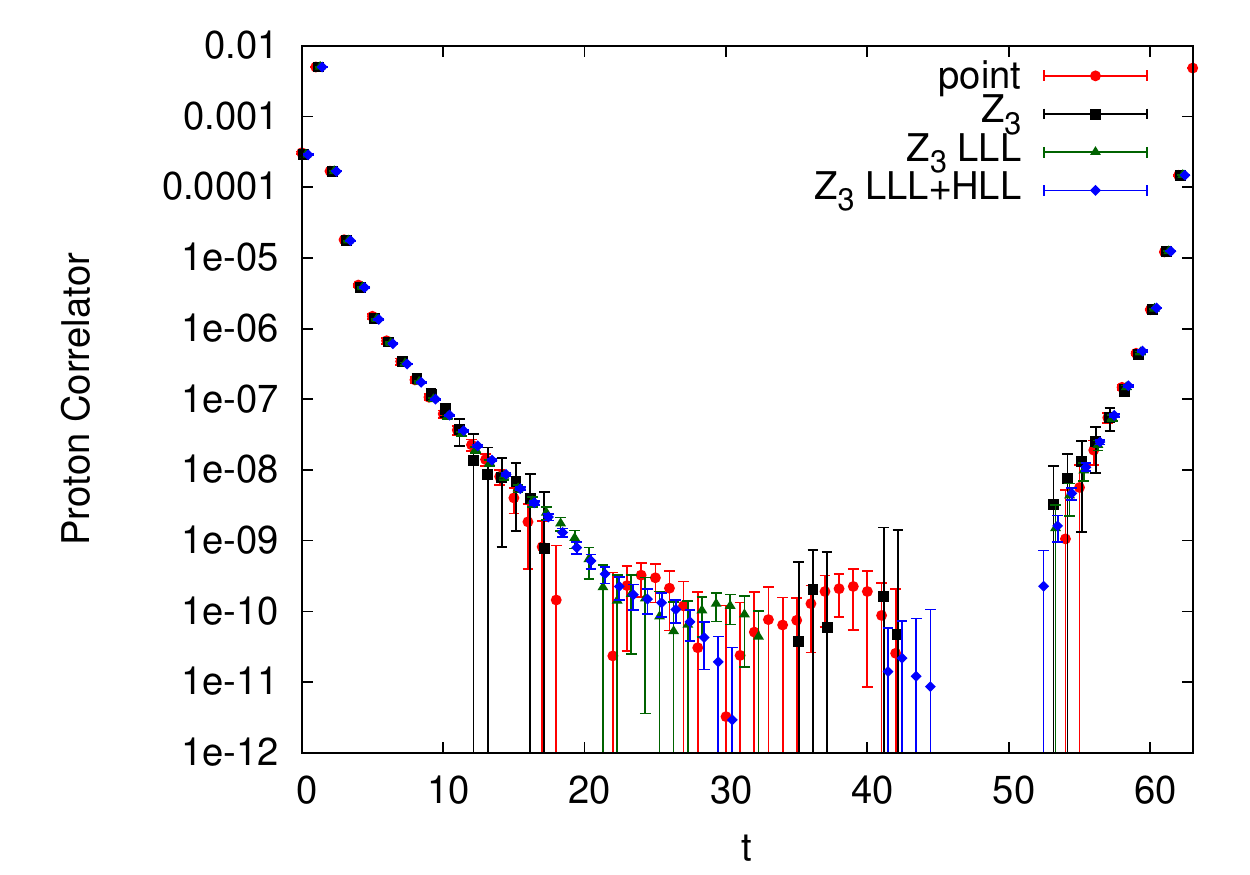}
\includegraphics[width=0.47\textwidth,height=0.28\textwidth,clip=true]{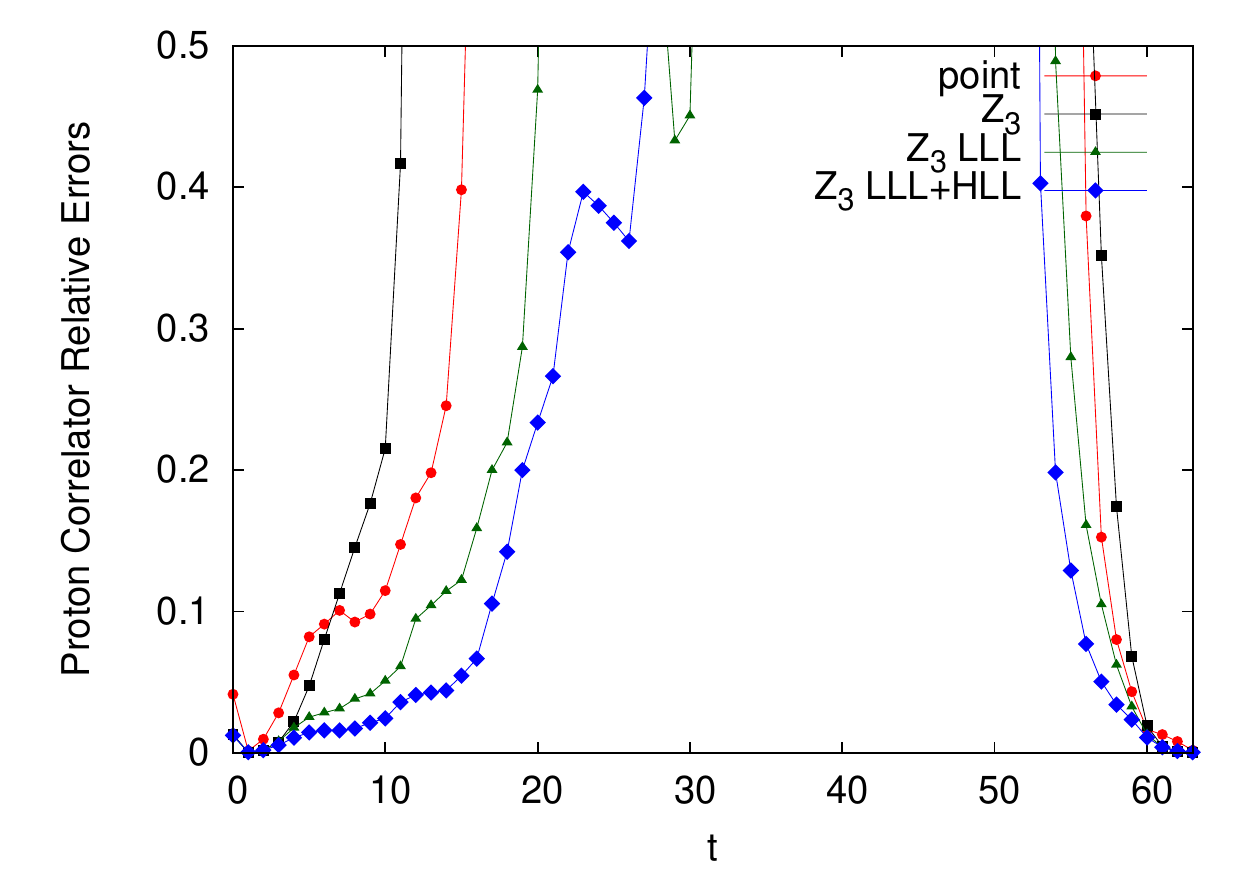}
\end{center}
\vspace{-0.3in}
\caption{The same as figure 2 for the nucleon correlator with $m_{\pi} \sim 300$ MeV.}
\vspace{-0.12in}
\end{figure}

\vspace{-0.15in}
In addition to the above speed up procedure we also substitute the
low-frequency part of the noise estimated correlator with the exact
one obtained from the eigenmodes, which has been shown to reduce the
%### [gs04 is undefined.]
variance~\cite{fjo05,kfh07}.  To do that we employ $Z_3$ noise
grid source with support on certain spatial grid points in a time
slice to calculate the quark propagator which amounts to a many-to-all
approach as opposed to the all-to-all approach.  
This reduces the noise contamination in the high frequency part of the correlators.
The details are given
in ref.~\cite{big-paper}. In Figs. 2 and 3 we have plotted the vector
(at $m_{\pi} \sim 200$ MeV) and nucleon (at $m_{\pi} \sim 300$ MeV)
correlators with point source, $Z_3$ grid source, and $Z_3$ grid
source with low mode substitution. On the right we also plot
the relative errors which clearly show that $Z_3$ grid source with
low mode substitution is the better option.
%%%%%%%%%%%%%%%%%%%%%%%%%%%%%%%%%%%%%%%%%%%%%%%%%%%%%%
\vspace{-0.1in}
\section{Numerical details}
\vspace{-0.1in}
We have calculated overlap quark propagators with overlap valence quarks by using 2 + 1 flavor domain
wall fermion gauge configurations generated by RBC and UKQCD collaborations~\cite{RBC-UKQCD08}. Three sets of lattices are used with the four-dimensional sizes of $16^3 \times 32, 24^3 \times 64 (a^{-1} =
1.73(3) \,\hbox{GeV})$, and $32^3 \times 64 (a^{-1} = 2.32(3)\, \hbox{GeV})$ with several sea quark masses each.
The Zolotarev rational polynomial approximation up to 14th degree is used to approximate
the matrix sign function and for the window [0.031, 2.5], the approximation to
the sign function is better than $3.3 \times 10^{-10}$~\cite{big-paper}. 
Furthermore, low-mode deflation is used for both the inner and outer inversion of the overlap operator. Utilizing HYP smearing the largest absolute 
values of the overlap  eigenvalues we deflate are 0.2, 0.125, and 0.22 on
$16^3\times 32, 24^3\times 64$, and $32^3\times 64$ lattices with 100, 400, and 200 eigenvectors, respectively. 
\vspace{-0.15in}
\section{Results}
\vspace{-0.1in}
The results reported here are for the $32^3 \times 64$ lattice with lattice spacing $a^{-1} = 2.32(3)$ GeV.
In Fig. 4(left) we plot $(m_{\pi}a)^2$ as a function of valence quark
masses $ma$ in the range [$0.0046 - 3.0$] and for three sea quark masses.
One can observe the 
nearly linear behavior up to a heavy quark mass. In Fig. 5(right) we also
plot $(m_{\pi}a)^2/ma$ vs $ma$ for the same range of
quark masses. One can observe the partially quenched chiral logarithm 
due to the mismatch of valence and sea quark masses. It is evident that
the divergence becomes prominent with the increase of sea quark mass when the valence quark masses are 
lighter than the sea. In the future we will 
estimate the chiral log
parameters for these sets of data.
\begin{figure}
\vspace{-0.2in}
\begin{center}
\includegraphics[width=0.47\textwidth,height=0.28\textwidth,clip=true]{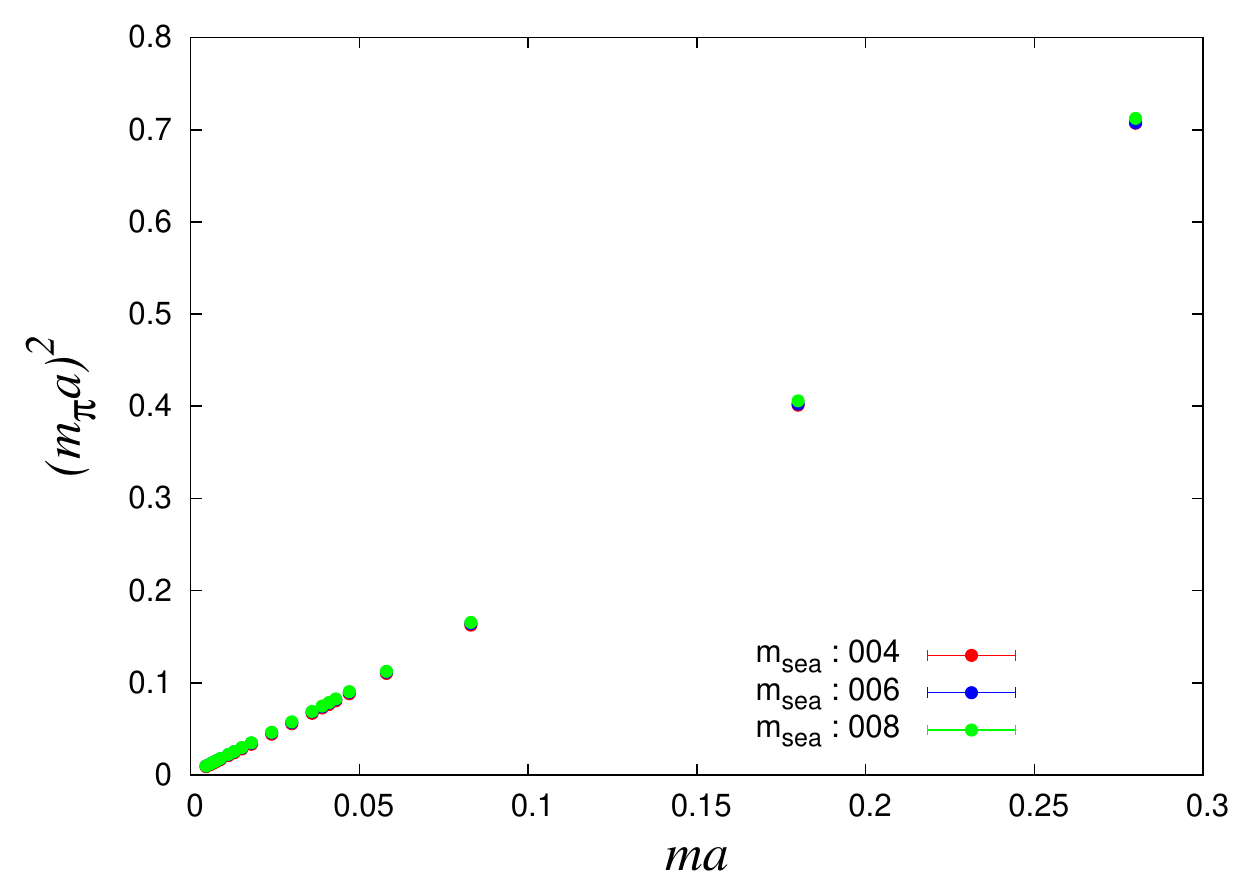}
\includegraphics[width=0.47\textwidth,height=0.28\textwidth,clip=true]{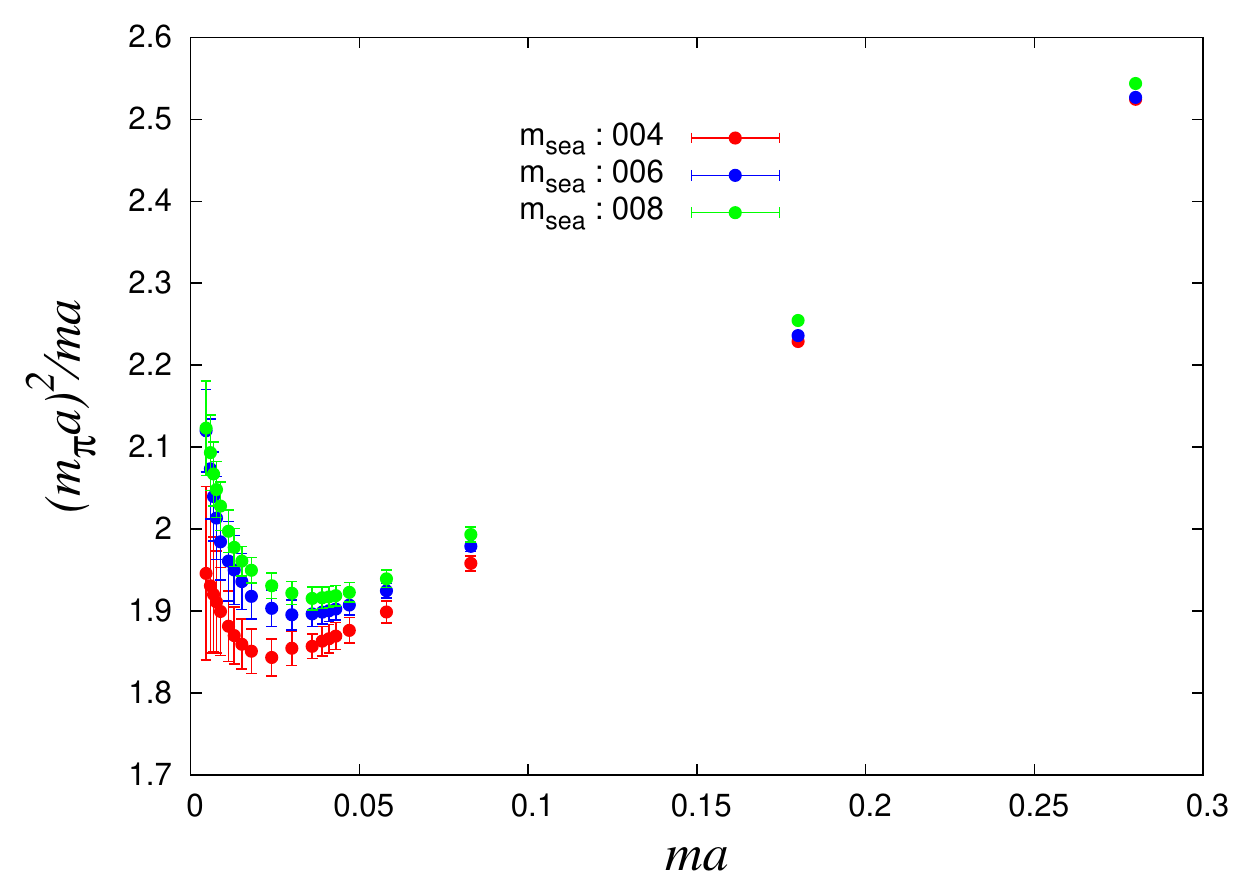}
\end{center}
\vspace{-0.18in}
\caption{(left) $(m_{\pi}a)^2$ is plotted as a function of $ma$ for the $32^3\times 64$ lattice with quark masses $ma$ = [0.0046 -- 0.3]. (right) $(m_{\pi}a)^2/ma$ vs $ma$ showing chiral logarithm for the same range of quark masses.}
\end{figure}

The effect of partial quenching is more evident in the case of the scalar 
meson. If the valence quark mass is lower than the sea mass then the lowest
state in the scalar meson spectrum will be the would-be $\eta\pi$ state
which has negative contribution  to the correlator. This has been
investigated previously by various groups~\cite{Bardeen:2001jm, Mathur:2006bs,pre06,ab06,wlo09}. 
In Fig. 6(left) we plot scalar
correlators for the sea mass 0.004 and for several valence quark
masses. It is evident that as one increases valence mass, effect of the
negative dip, which is a signature of the would-be $\eta\pi$ state,
reduces. In Fig. 6(right) we plot scalar correlators for a given valence
quark mass and for three sea quark masses. It is also evident that as 
one decreases the sea mass the contribution to the partially quenched ghost reduces.
\newpage
\begin{figure}
\vspace{0.1in}
\begin{center}
\includegraphics[width=0.47\textwidth,height=0.28\textwidth,clip=true]{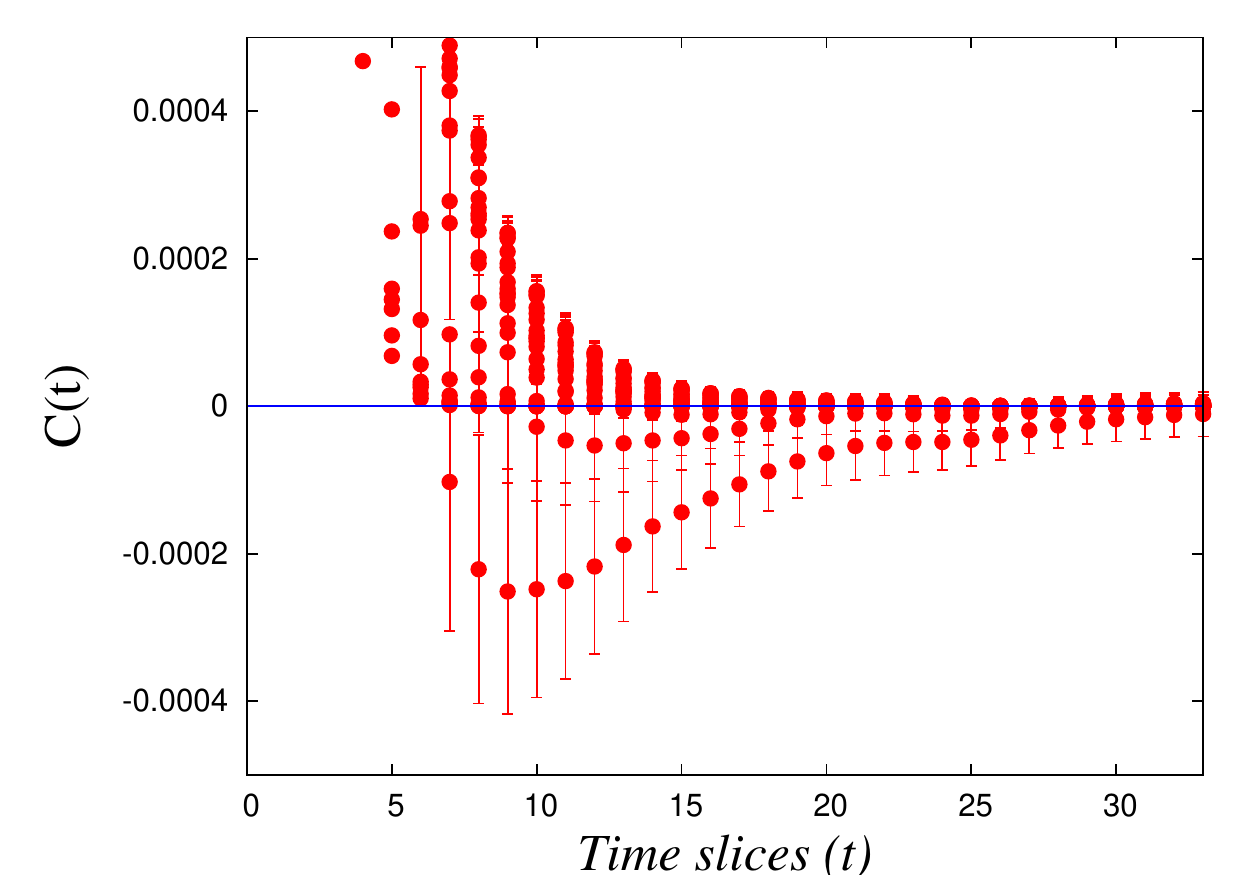}
\includegraphics[width=0.47\textwidth,height=0.28\textwidth,clip=true]{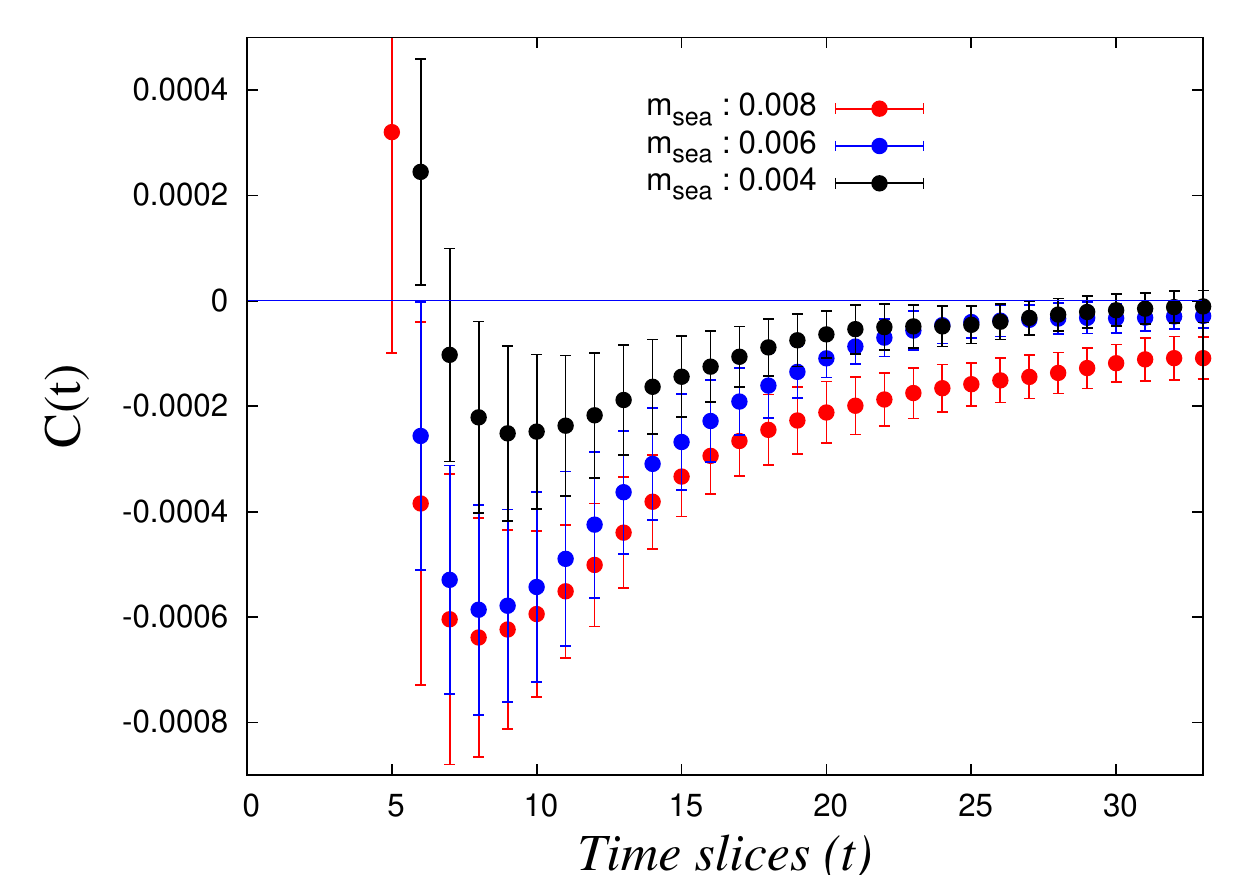}%\\
%\hspace{0.1in}(a)\hspace{2.8in}(b)
\end{center}
\vspace{-0.15in}
\caption{(left) Isovector scalar meson correlators are plotted for various valence quark masses and for $m_{s} = 0.004$. (right)The same correlators at a fixed valence mass (lower than smallest $m_{s}$) and for various $m_{s}$.}
\vspace{-0.1in}
\end{figure}

If $m_{vv'}/m_{ss'}$ is the mass of the pseudoscalar meson made up of the valence/sea quark and antiquark and $m_{vs}$ is the mass of the mixed valence and sea pseudoscalar meson, then the leading order (LO) pseudoscalar meson masses are given as
\vspace{-0.1in}
\begin{eqnarray}   \label{mixed_mass_relation}
m_{vv'}^2 &=& B_{ov}(m_v +m_{v'}), \nonumber \\
m_{vs}^2 &=& B_{ov}m_v + B_{dw}(m_s + m_{res}) + a^2\Delta_{mix}, \nonumber \\
m_{ss'}^2 &=& B_{dw}(m_s + m_{s'} + 2 m_{res}),
\vspace{-0.15in}
\end{eqnarray}
where $\Delta_{mix}$ is a low energy constant representing $\mathcal{O}(a^2)$ discretization 
dependence in the LO mixed-action chiral Lagrangian. $\Delta_{mix}$ is also a measure of unitarity violation due to mixed action formalism at finite lattice spacing, which should vanish at the continuum limit. One can notice that $\Delta_{mix}$ term enters only in $m_{vs}$. We have measured $\Delta_{mix}$ term in ref.~\cite{big-paper} and have found that it is $\sim$ 7 times smaller
than the case of DWF valence on staggered sea~\cite{ow08} and $\sim$ 18 times smaller than that of overlap
on Wilson sea~\cite{dfh07}. For a 300 MeV pion on the
$24^3 \times 64/32^3 \times 64$ lattice with $a \sim 0.12/0.085$ fm, the shift in mass due to
$\Delta_{mix}$ is $\sim 19/10$ MeV. 
This small shift due to $\Delta_{mix}$ could admit a better extrapolation 
with the help of MAPQ$\chi$PT.

In Fig. 7 we show meson masses obtained in a large range of valence
quark masses at the lightest sea mass, $m_{sea} = 0.004$. On the left side,
we have plotted pseudoscalar, vector, axial and scalar masses as a function of
$(m_{\pi}a)^2$. The threshold decay channels are also depicted with solid lines to 
indicate two-meson energy. 
For the vector channel, as one can observe, our data points are much
below the threshold $\pi\pi$ P-wave energy indicating that $\pi\pi$ state 
lies higher than the $\rho$ on on this lattice 
for the given range of quark masses. However, the axial channel decay 
threshold is overlapping our data which indicates that our data either
represents the axial resonance ($a_1$) or the two particle state $\rho\pi$. 
If this is indeed the $\rho\pi$ state, then this will be the first 
observation of this state on the lattice. However, to conclude that one 
needs to show the existence of it along with the axial vector meson, 
and furthermore, the required volume dependence of a two particle state. 
On the right hand side we have plotted the meson masses in the charm quark mass range. 
In the future, we will do chiral extrapolation with MAPQ$\chi$PT and then continuum extrapolation to obtain physical values for these meson masses.
\begin{figure}
\vspace{-0.1in}
\begin{center}
\includegraphics[width=0.47\textwidth,height=0.28\textwidth,clip=true]{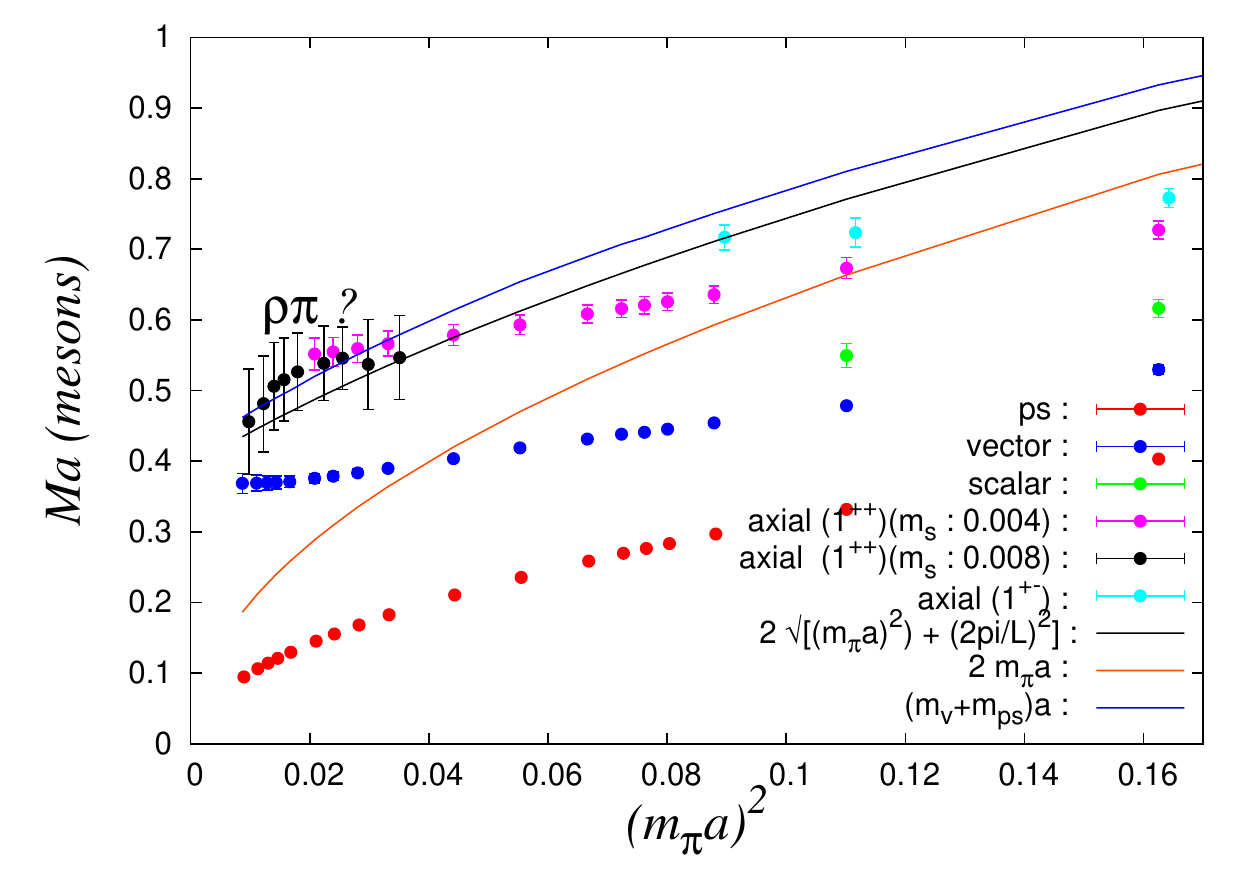}
\includegraphics[width=0.47\textwidth,height=0.28\textwidth,clip=true]{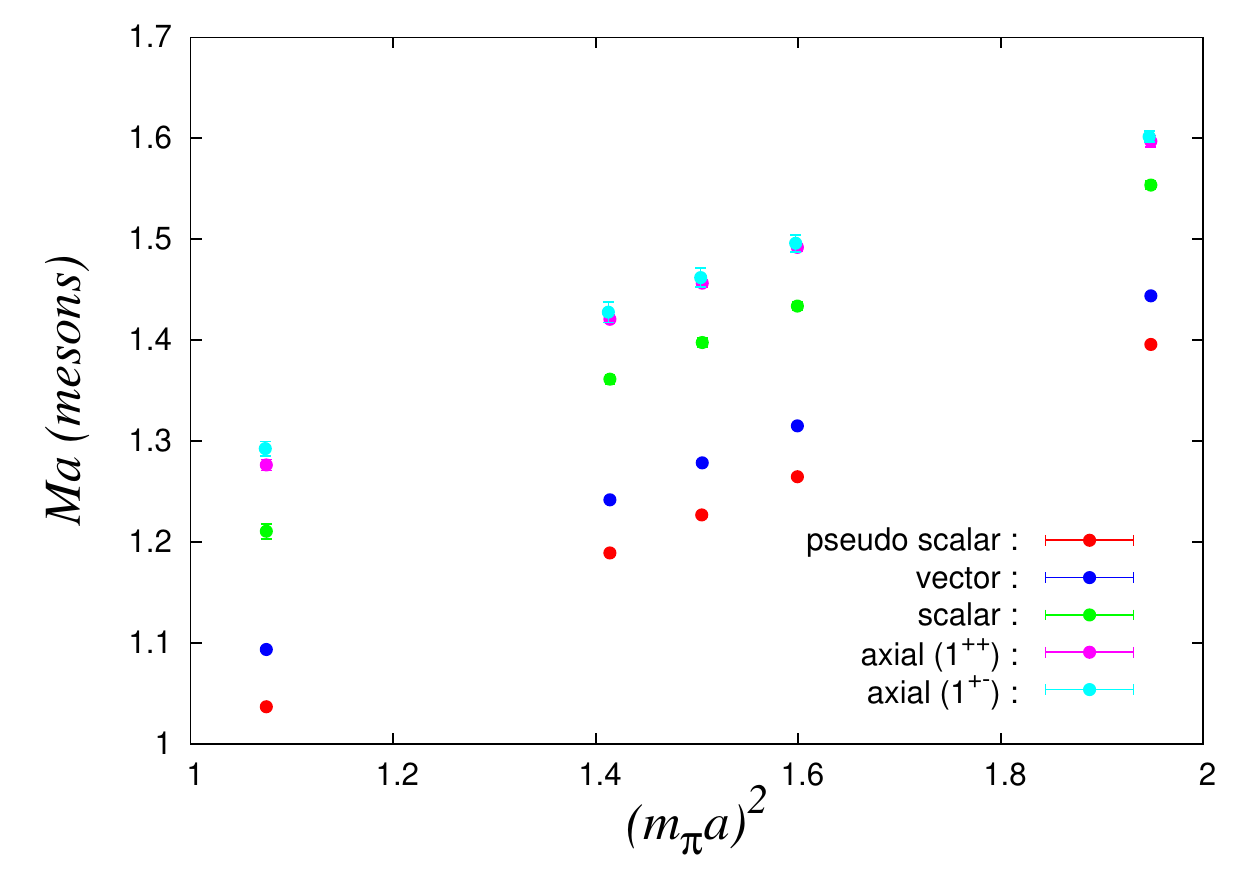}
\end{center}
\vspace{-0.2in}
\caption{(left) Light quark meson masses are plotted over a range of $(m_{\pi}a)^2$. Threshold decay channels are also indicated by solid lines. (right) Meson masses in the charm quark mass ranges.}
\vspace{0.05in}
\end{figure}
\begin{figure}
\begin{center}
\includegraphics[width=0.47\textwidth,height=0.28\textwidth,clip=true]{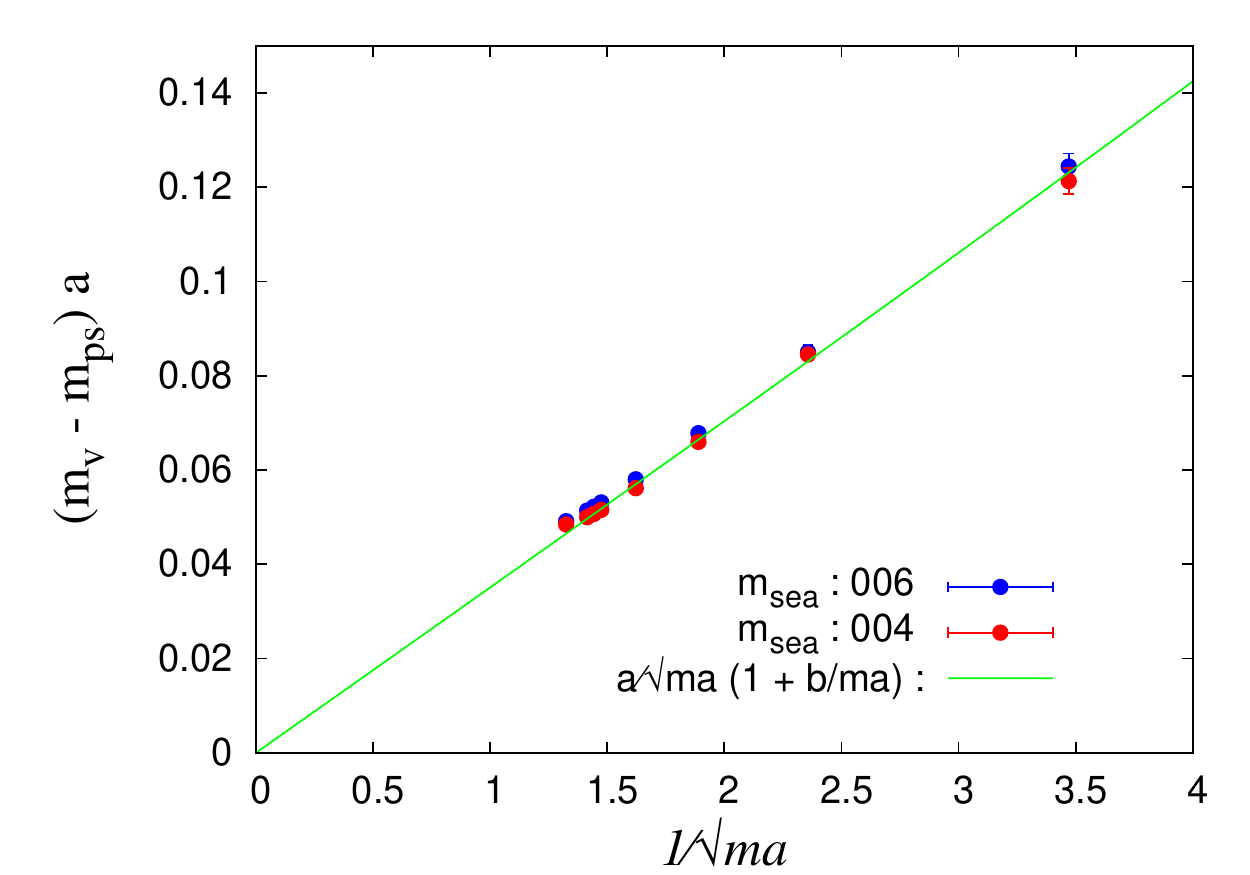}
\includegraphics[width=0.47\textwidth,height=0.28\textwidth,clip=true]{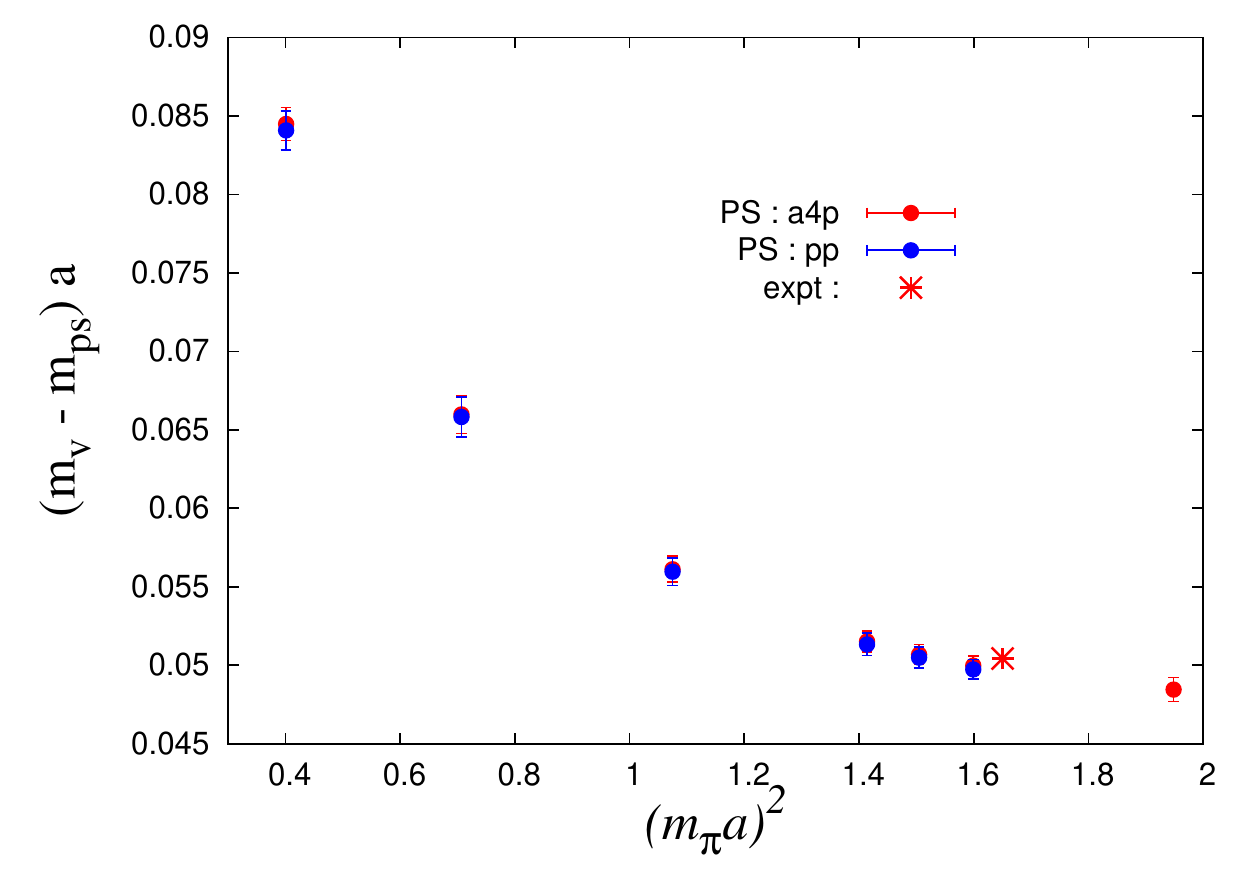}
\end{center}
\vspace{-0.15in}
\caption{(left) The hyperfine splitting is plotted in the charm quark mass region as a function of $\frac{1}{\sqrt{ma}}$ showing its linear dependence. (right) The same as a function of
$(m_{\pi}a)^2$ along with experimental data.}
\vspace{-0.1in}
\end{figure}

In Fig.8 we plot the hyperfine splitting (HFS) between heavy vector and pseudoscalar meson masses. 
Over the years, HFS obtained by lattice calculations are smaller than its experimental value. It was argued that lattice spacing errors along with dynamical fermion effects and non-inclusion of the disconnected insertion contributed to this discrepancy. The HFS is expected to scale like $\propto \frac{1}{\sqrt{m}}$
%%
%\begin{equation}
%\Delta E_{HFS} \propto \frac{1}{\sqrt{m}}.
%\end{equation}
%%
to leading order in $m$. Based on this expectation, in Fig.8(left) we plot the HFS obtained in the charm quark mass range for two sea quark masses. It is to be noted that the data points are linear in this mass range and one can expect a smaller $m^2$ dependence.
We fit the HFS in this range to the form
%
%\begin{equation}
$\Delta E_{HFS} = \frac{a}{\sqrt{m}}(1 + \frac{b}{m})$,
%\end{equation}
%
which includes the next term in large $m$ expansion. The fit value obtained are $a = 0.0349(5)$ and  $b = 0.0014(2268)$ which shows that the $1/\sqrt{m}$ dependence is pretty accurate.
In Fig. 8(right) we plot the HFS as a function of $m_{\pi}^2a^2$ along with the experimental number
and one can observe that the experimental number is very close to one of our data point. Of course one needs to do continuum and chiral extrapolation to obtain its final value. This is very encouraging in the sense that other charm physics can also be obtained from this set of lattice.
\newpage
\vspace*{-0.5in}
\section{Summary and outlooks}
\vspace{-0.1in}
Overlap valence on $2 + 1$ flavor DWF configurations is a promising
approach to do lattice QCD simulation with light, strange, and charm
quark together in the same fermion lattice formulation.  We have
demonstrated that the eigenvalue deflation along with HYP smearing is
very efficient procedure for inversion and the the $Z_3$ grid source
with low-mode substitution can reduce error in two-point correlator up
to a factor 4. Results shown for the meson spectrum are very
encouraging. With appropriate MAPQ$\chi$PT we will next do the chiral
extrapolation and then continuum extrapolation. Furthermore, we found
that the $1/\sqrt{m}$ behavior in the HFS extends to the charm quark
range and the result for the HFS is quite encouraging. This motivates
us to study charm-light, charm-strange and charm-charm spectrum and
other related phenomenologies in the future.
\vspace{-0.05in}
\section{Acknowledgments}
\vspace{-0.05in}
This work is partially support by U.S. DOE Grants No. DE-FG05-84ER40154, No. DEFG02-
95ER40907 and No. DE-FG02-05ER41368.
The research of N.M. is supported under grant No. DST-SR/S2/RJN-19/2007, India.

\end{document}